\documentclass[reprint, amsmath, amssymb, aps]{revtex4-2}

\usepackage{graphicx}
\usepackage{dcolumn}
\usepackage{bm}
\usepackage{xcolor}

\usepackage{braket}
\usepackage{amsmath}
\usepackage{amssymb}

\usepackage{xurl}

\usepackage{algorithm}
\usepackage{algpseudocode}

\begin{document}

\preprint{APS/123-QED}

\title{A Time Optimization Framework for the Implementation of Robust and Low-latency Quantum Circuits}

\author{Eduardo Willwock Lussi}
 \altaffiliation[Also at ]{Quantuloop.}
 \email{eduardolussi@gmail.com}
\author{Rafael de Santiago}
 \email{r.santiago@ufsc.br}
\affiliation{Informatics and Statistics Department, Federal University of Santa Catarina \\
             Florianópolis, Santa Catarina, Brazil}

\author{Eduardo Inacio Duzzioni}
 \email{duzzioni@gmail.com}
\affiliation{Department of Physics, Federal University of Santa Catarina \\
             Florianópolis, Santa Catarina, Brazil}

\date{\today}

\begin{abstract}
Quantum computing has garnered attention for its potential to solve complex computational problems with considerable speedup. Despite notable advancements in the field, achieving meaningful scalability and noise control in quantum hardware remains challenging. Incoherent errors caused by decoherence restrict the total computation time, making it very short. While hardware advancements continue to progress, quantum software specialists seek to minimize quantum circuit latency to mitigate dissipation. However, at the pulse level, fast quantum gates often lead to leakage, leaving minimal room for further optimization. Recent advancements have shown the effectiveness of quantum control techniques in generating quantum gates robust to coherent error sources. Nevertheless, these techniques come with a trade-off -- extended gate durations. In this paper, we introduce an alternative pulse scheduling approach that enables the use of both fast and robust quantum gates within the same quantum circuit. The time-optimization framework models the quantum circuit as a dependency graph, implements the fastest quantum gates on the critical path, and uses idle periods outside the critical path to optimally implement longer, more robust gates from the gate set, without increasing latency. Experiments conducted on IBMQ Brisbane show that this approach improves the absolute success probability of quantum circuit execution by more than 25\%, with performance gains scaling as the number of qubits increases.
\end{abstract}


\maketitle

\section{\label{sec:introduction}Introduction}

Quantum computing has gained recognition for its potential to solve complex problems exponentially faster than classical computers. The field has evolved from theoretical quantum algorithms \cite{DeutschJozsa, Grover, Shor, BernsteinVazirani} to the first demonstrations using early-stage quantum computers in the early 2000s \cite{Chuang1998, knill2000algorithmic, vandersypen2001experimental}. Commercial products entered the scene in 2007 when D-Wave Systems unveiled a 16-qubit quantum annealing computer \cite{dwavedemo2007}. IBM also made significant contributions to the field, launching the IBM Quantum Experience in 2016, which provided cloud access to the first quantum computer \cite{PRNewswire2016}. IBM's latest quantum computer, the Condor, now boasts 1,121 qubits \cite{castelvecchi2023ibm}, intending to reach 100,000 qubits by 2033 \cite{Gambetta2023}.

While quantum computing has attracted significant attention, there are potential challenges and limitations in achieving meaningful scalability and noise control. Quantum errors can be classified into two categories \cite{Wallman2015}. Coherent errors are represented by unitary operations, behaving like extra quantum gates. They are referred to as ``coherent'' because they do not mix the quantum state. Examples include leakage, dephasing, cross-talk, and calibration errors. Incoherent errors are irreversible and cannot be described by unitary operations. They are modeled as mixed states, equivalent to the system undergoing a quantum operation with a given probability $p$ \cite[Section 2.4.1]{NielsenChuang}. The most significant incoherent errors in this context are those caused by energy dissipation, which gradually causes the quantum state to be lost to the environment. This limits the effective computation time on current quantum computers. Energy dissipation is characterized by two parameters: $T_1$, the amplitude damping, which describes the decay of the state $\ket{1}$ to $\ket{0}$; and $T_2$, the phase damping, which determines the time over which the qubit retains phase information (the $xy$ plane).

To address quantum errors, there are three primary approaches. Quantum Error Correction (QEC) operates at a higher abstraction level by encoding logical qubits using multiple physical qubits. This redundancy allows the system to reconstruct the logical state even if some physical qubits fail \cite{ShorQEC}, \cite[Chapter 10]{NielsenChuang}. Quantum Error Mitigation (QEM) focuses on estimating ideal expectation values by post-processing noisy measurement data \cite{QEM}. For instance, the zero-noise extrapolation method \cite{ZeroNoiseExtrapolation} artificially amplifies errors in a controlled manner, enabling an estimate of the error-free execution. Lastly, Quantum Error Suppression (QES) works at the hardware level, aiming to prevent errors before they occur. Superconducting qubits are controlled via microwave pulses, and quantum gates are implemented by modulating pulse attributes such as frequency, amplitude, and phase. The correct manipulation of these attributes, combined with other interventions at this hardware level, characterizes error suppression.

Recently, Q-CTRL, an Australian company specializing in quantum control solutions, has made significant advancements in the field by demonstrating the effectiveness of quantum control techniques in improving the quality of quantum algorithms through the generation of noise-robust and high-fidelity quantum gates \cite{Carvalho2021, Baum2021, Mundada2023, QCTRLIBMQ}. Integrated into the Boulder Opal tool \cite{QCTRLBoulderOpal, QCTRLDesign}, Carvalho et al. \cite{Carvalho2021} present an optimization-based technique that incorporates error processes into the control pulse Hamiltonian, generating pulses that are robust to these errors. The study specifically targets the coherent error processes of dephasing and pulse amplitude errors. However, these techniques often require extended gate durations, which pose limitations when considering energy dissipation errors.

Reducing circuit latency is an important aspect in both quantum control \cite{Baum2021, Semola, Shi2019} and quantum compilation \cite{Gokhale2020, Murali2019, Lao2022}. To generate faster quantum gates, stronger control pulses are required, but this strength can excite states beyond $\ket{1}$, resulting in leakage errors. To suppress this problem, techniques such as Derivative Removal by Adiabatic Gate (DRAG) are commonly employed \cite{DRAG, FASTDRAG}. These techniques reduce high-frequency components in the control signals, enabling the generation of faster control pulses without inducing leakage. Additionally, machine learning approaches have also proven useful for generating fast quantum gates \cite{Baum2021, Semola}. In experiments conducted on IBMQ, the Q-CTRL autonomous agent \cite{Baum2021, QCTRLCalibration} generated quantum gates three times faster than the default DRAG used for single-qubit gates, without additional leakage. Because the learning process took several hours, the agent was also able to adapt to calibration drifts and sustain performance over several days.

Here lies the challenge: incorporating robustness into fast control pulses is inherently difficult because these pulses generally need to minimize leakage errors and there is no space for further optimization. Conversely, longer control pulses can achieve greater robustness, but their extended duration increases susceptibility to energy dissipation in NISQ quantum computers.

In this paper, we present a framework that enables the use of both fast control pulses and long noise-robust control pulses within the same quantum circuit without additional latency. Considering a gate set with multiple implementations of the same quantum gate (e.g., fast gates and longer, robust gates), the algorithm models the circuit as a quantum operation dependency graph and uses project management techniques to select the optimal gate implementations. To maintain low latency without sacrificing robustness, the algorithm places fast control pulses on the critical path and uses idle periods to implement longer control pulses, since latency is not compromised.

Randomized benchmarking experiments on the IBMQ Brisbane quantum computer demonstrate that this technique can improve the success probability of quantum circuit execution by more than 25\%. The most significant improvements appear in circuits with more qubits, which generally have more idle periods.

This paper is organized as follows. Section \ref{sec:pulse} introduces pulse-level programming, including quantum gate calibration and important aspects of pulse shaping. Section \ref{sec:compilation} presents key techniques in quantum circuit compilation. Section \ref{sec:cpm} covers the Critical Path Method used by the time-optimization algorithm. Section \ref{sec:proposal} details the optimization framework and algorithm, including complexity considerations. Section \ref{sec:exp} reports the experiments validating the approach, and Section \ref{sec:conclusion} presents conclusions and future work.

\section{\label{sec:pulse}Pulse-level Programming}

Quantum gates in superconducting quantum computers are implemented through microwave pulses that, at the appropriate frequency, induce oscillations between the $\ket{0}$ and $\ket{1}$ states, known as Rabi oscillations. To implement the desired operation, these oscillations are controlled in time by modulating the pulse attributes.

A pulse is characterized by a waveform, typically represented by sine and cosine functions. In this context, $IQ$ modulation is often employed, where the in-phase ($I$) and quadrature ($Q$) components correspond to two distinct pulses with a phase shift of $\pi/2$. The final pulse is obtained by combining the $I$ and $Q$ components, resulting in a pulse with an amplitude and phase dependent on these components. Specifically, the pulse can be described as: 
\begin{equation} 
\begin{aligned} 
    I(t) &= A_I(t)\cos{\left(\omega t + \phi\right)}, \\ 
    Q(t) &= A_Q(t)\sin{\left(\omega t + \phi\right)}, 
\end{aligned} 
\end{equation} 
where $\omega$ is the pulse frequency, $A(t)$ is the dimensionless amplitude modulation, $\phi$ is the phase, and $t$ denotes time. The control Hamiltonian is given by ($\hbar=1$):
\begin{equation}
    H(t)=\frac{1}{2}\left(I(t)\sigma_x + Q(t)\sigma_y\right).
\end{equation}

It becomes clear that the $I$ component induces oscillations around the $x$ axis, while the $Q$ component induces oscillations around the $y$ axis. Intermediate phase values, which can also be modulated by $\phi$, cause oscillations along other axes within the $xy$ plane. The frequency $\omega$ should correspond to the qubit's resonance frequency, inducing $\ket{0}\leftrightarrow\ket{1}$ transitions. The amplitude $A$ controls the rate of oscillation: a higher amplitude results in faster oscillations. 

\begin{table*}
    \centering
    \begin{tabular}{c|c|c}
        \textbf{Shape} & \textbf{Parameters} & \textbf{Function} \\
        \hline
        Square & 
        \(\begin{array}{r@{\,:\,}l}
            A & \text{amplitude} \\ 
            d & \text{duration}
        \end{array}\) & 
        $f(t)=A$
        \\
        \hline
        Gaussian & 
        \(\begin{array}{r@{\,:\,}l}
            A & \text{amplitude} \\ 
            d & \text{duration} \\
            \sigma & \text{std. deviation} \\
        \end{array}\) & 
        \(\begin{aligned}
            f(t) &= A \times \mathcal{N}(g(t)) \text{ for } \mu=d/2
        \end{aligned}\)
        \\
        \hline
        Gaussian-Square & 
        \(\begin{array}{r@{\,:\,}l}
            A & \text{amplitude} \\ 
            d & \text{duration} \\
            \sigma & \text{std. deviation} \\
            w & \text{square portion width} \\
            r & \text{risefall factor ($d-w$)}
        \end{array}\) & 
        \(\begin{aligned}
            h(t) &= 
            \begin{cases}
                g(t) \text{ for } \mu=r   & \text{if $t < r$}          \\
                1                         & \text{if $r \leq t < r+w$} \\
                g(t) \text{ for } \mu=r+w & \text{if $r + w \leq t$}   \\
            \end{cases} \\
            f(t) &= A \times \mathcal{N}(h(t))
        \end{aligned}\)
        \\
        \hline
        DRAG & 
        \(\begin{array}{r@{\,:\,}l}
            A & \text{amplitude} \\ 
            d & \text{duration} \\
            \sigma & \text{std. deviation} \\
            \beta & \text{complex amp. strength} \\
        \end{array}\) & 
        \(\begin{aligned}
            f_I(t) &= A \times \mathcal{N}(g(t)) \text{ for } \mu=d/2 \\
            f_Q(t) &= f_I'(t) = -\frac{t-d/2}{\sigma^2} \times f_I(t) \\
            f(t) &= f_I(t) + i \times \beta \times f_Q(t) \\
        \end{aligned}\)
        \\
    \end{tabular}
    \caption{Summary of common pulse shapes used in the implementation of quantum gates.}
    \label{tab:pulse-shape}
\end{table*}

\subsection{\label{subsec:rabi-exp}Rabi Experiment}

To implement a control pulse, we first need to understand the relationship $(I, Q) \leftrightarrow (A_I, A_Q)$—that is, we need to determine the hardware input amplitude values $A_I$ and $A_Q$ corresponding to the Rabi oscillations $I$ and $Q$. The Rabi experiment is a characterization procedure that involves selecting a set of amplitude values and measuring the corresponding oscillation frequencies. The Rabi oscillations for each pulse amplitude can be described by:
\begin{equation}\label{eq:rabi-oscillation} 
    y(t)=\mathcal{A}_y\cos^2{\left(2\pi \Omega t + \Phi_y\right)} + \delta_y, 
\end{equation} 
where $y(t)$ is the measured signal (e.g., normalized $z$-axis projection), $t$ is time, $\Omega$ is the Rabi frequency, and $\mathcal{A}_y$, $\Phi_y$, and $\delta_y$ are fitting parameters \cite{QCTRLCalibration}.

By fitting the experimental data to equation \ref{eq:rabi-oscillation}, the Rabi frequencies $\Omega$ for each pulse amplitude can be determined. The relationship between pulse amplitude and Rabi frequency is typically linear and symmetric for positive and negative amplitudes. To create a continuous function for translating between $\Omega$ and $A$ and encompass intermediate values from the set of amplitudes, the experimental data is interpolated. 

Evidently, this interpolation does not represent the ideal $\Omega \leftrightarrow A$. Specifically, we only know the exact Rabi frequencies for the set of amplitudes selected for the Rabi experiment, so estimating other amplitude values may lead to under- or over-rotations. To refine this translation, the control pulse must be fine-tuned. This process typically involves sweeping the amplitude around the values estimated by the interpolation and identifying the amplitude that achieves the highest fidelity.

\subsection{Pulse Shaping}

Control pulses are usually described in discrete time intervals, known as the sampling time $dt$. A complete control pulse thus consists of a sequence of pulse amplitudes with a duration of $dt$.

\begin{figure}[h]
    \centering
    \includegraphics[width=1\linewidth]{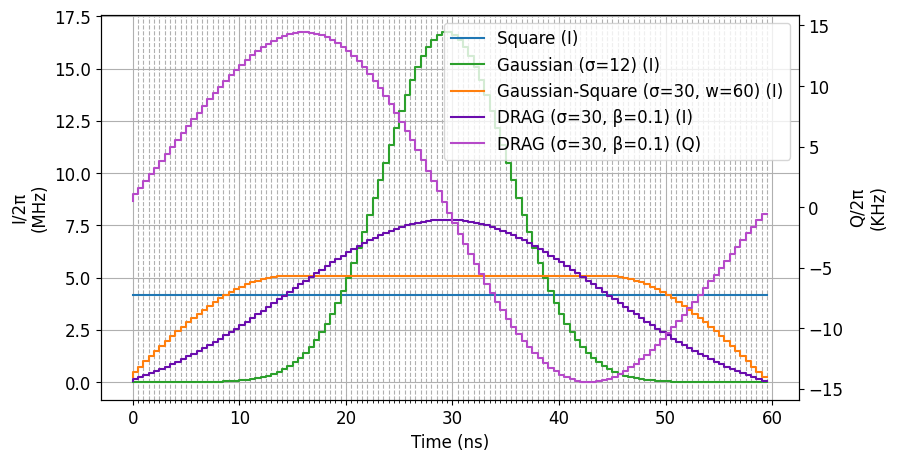}
    \caption{Square, Gaussian ($\sigma=12$), Gaussian-Square ($\sigma=30$, $w=60$) and DRAG ($\sigma=30$, $\beta=0.1$) pulse shapes for the implementation of a $\pi/2$ rotation.}
    \label{fig:pulse-shape}
\end{figure}

Several standard pulse shapes are commonly used to implement quantum gates. Using the normalization function: 
\begin{equation} 
    \mathcal{N}(f(t))=\frac{f(t)-f(-1)}{1-f(-1)}, \text{ for } 0 \leq t \leq d
\end{equation}
where $d$ is the pulse duration, and the Gaussian function: 
\begin{equation} 
g(t) = \exp{\left(-\frac{(t-\mu)^2}{2\sigma^2}\right)}, 
\end{equation} 
the Square, Gaussian, Gaussian-Square, and DRAG pulse shapes are summarized in Table \ref{tab:pulse-shape}. Figure \ref{fig:pulse-shape} presents these control pulses for implementing a $\pi/2$ rotation (Sx or $\sqrt{X}$ gate).

The square pulse shape is characterized by a fixed amplitude. While straightforward to compute, the abrupt rise and fall of the amplitude can introduce errors due to the physical limitations of control equipment, which cannot instantaneously change the amplitude. The Fourier transform of a square pulse requires many high-frequency components, making precise control of amplitude and phase challenging \cite{Yan2019}. Moreover, these high-frequency components can induce leakage errors \cite{Yan2019}, where states beyond $\ket{1}$ are excited. However, the constant and typically low amplitude of this pulse can be advantageous in certain applications \cite{QFALussi}.

The other pulse shapes discussed here are generally preferred for quantum gate implementation. The Gaussian pulse has a smooth shape that avoids high-frequency components, though high amplitudes may be required depending on standard deviation $\sigma$ and especially in faster pulses, potentially introducing errors \cite{DRAG}.

The Gaussian-Square shape combines the smooth rise and fall of a Gaussian pulse with a central square portion defined by the width parameter $w$. This shape is useful when lower amplitude values are needed compared to a purely Gaussian shape, while still maintaining smoothness, as shown in Figure \ref{fig:pulse-shape}.

The Derivative Removal by Adiabatic Gate (DRAG) pulse, introduced in \cite{DRAG}, is designed to mitigate leakage errors. This approach combines a standard Gaussian pulse in the $I$ component with an additional derivative component in the $Q$ component. The derivative component smooths the original Gaussian shape by adjusting its phase, effectively filtering out unwanted frequency components. Higher-order derivatives can also be applied to both the $I$ and $Q$ components, as demonstrated effectively in \cite{FASTDRAG}. 

Other methods for pulse generation have also been proposed. \cite{QCTRLAndre, QCTRLDesign} describes an optimization-based strategy that incorporates the coherent error processes of dephasing and amplitude fluctuations into the Hamiltonian to generate waveforms robust to these errors. However, a significant limitation of these methods is that the optimization often requires extended gate durations. Machine learning approaches have also been explored \cite{QCTRLBaum, Semola}. The challenge posed by the arbitrary nature of quantum errors makes their characterization and mitigation quite difficult. Machine learning algorithms can learn to suppress these errors without prior knowledge of the quantum hardware. The rise of cloud-based quantum computing platforms, which execute billions of quantum circuits daily \cite{IBMQCircuitsDay}, offers a promising avenue for the development of these learning-based techniques.

In addition to quantum control, higher-level tasks are also important for the successful implementation of quantum applications. Several strategies are being developed in the field of quantum compilation, which will be described next.

\section{\label{sec:compilation}Quantum Circuit Compilation}

To execute a quantum circuit on quantum hardware, several compilation steps are necessary to translate the circuit into a sequence of pulses for the qubits. The default strategy used by most quantum computing platforms is referred to as Static Pulse Scheduling. This approach involves calibrating a universal gate set to perform well across all scenarios. The quantum compilation process mainly involves two tasks: (i) mapping the circuit's qubits onto the hardware topology and (ii) decomposing high-level quantum gates into the hardware’s gate set.

Many optimizations can be applied within this static approach. At the qubit mapping level (i), \cite{Lao2022} addresses environments where drive and measurement channels are shared between qubits. In such cases, qubits not involved in CZ operations that share a drive channel need to be detuned for the operation. To handle these constraints, they model the hardware mapping as a polynomial complexity optimization problem, aiming to minimize circuit latency while accounting for these classical constraints. Additional information can also be used in quantum compilation. For example, \cite{Murali2019} maps hardware qubits based on calibration data, considering the impact of CNOT and readout errors, and optimizes this mapping to minimize SWAP operations, which depend on the hardware topology and the quantum circuit.

There are also several strategies for optimizing quantum gate decomposition (ii). For instance, \cite{evandroDecomposition2024} reduces the number of CNOTs to at most $12n$, where $n$ is the number of control qubits, in $\mathcal{O}(n)$ time. Their evaluation shows a reduction in the number of CNOTs from 101,245 to 2,684 in a 114-qubit Grover's algorithm. At the gate level, \cite{Maldonado2022} optimizes the parameters $\theta$, $\phi$, and $\lambda$ of U3 gates based on the quantum computer's decoherence times.

In terms of gate set calibration, some common techniques are often employed. For example, IBMQ typically uses the gate set $\left\{\text{Sx}, \text{Rz}, \text{CNOT}\right\}$ for good reasons. Consider the single-qubit U3 gate described below, parameterized by three Euler angles ($\theta$, $\phi$, and $\lambda$): 
\begin{equation} 
    \text{U3}\left(\theta,\phi,\lambda\right)= 
    \begin{bmatrix} \cos{\frac{\theta}{2}} & -e^{i\lambda}\sin{\frac{\theta}{2}} \\      
           e^{i\phi}\sin{\frac{\theta}{2}} &  e^{i\left(\phi+\lambda\right)}\cos{\frac{\theta}{2}} 
    \end{bmatrix}. 
    \end{equation} 
By calibrating an Sx gate for each qubit, any single-qubit gate can be implemented using the following decomposition \cite{McKay2017, U3Qiskit}: 
\begin{equation} 
    \label{eq:static-decomposition}
    \text{U3}\left(\theta, \phi, \lambda\right)= 
    \text{Rz}\left(\lambda\right) 
    \text{Sx}
    \text{Rz}\left(\theta\right) 
    \text{Sx}^{-1} 
    \text{Rz}\left(\phi\right), 
\end{equation} 
where the Rz gate is implemented virtually by shifting the phase of all subsequent pulses affecting the qubit \cite{McKay2017}.

It is also possible to reduce the number of quantum gates in this decomposition if arbitrary $x$ rotations are feasible, using the following decomposition:
\begin{equation}
    \label{eq:dynamic-decomposition}
    \text{U3}\left(\theta, \phi, \lambda\right)=
    \text{Rz}\left(\lambda-\pi/2\right)
    \text{Rx}\left(\theta\right)
    \text{Rz}\left(\phi-3\pi/2\right).
\end{equation}
In fact, \cite{Gokhale2020} proposed scaling the default pulse calibration for the Sx gate to perform the Rx gate, thus requiring no additional calibration. This same approach was also used to reduce multi-qubit gate duration, particularly with the Cross-Resonance gate, which typically implements a CNOT gate, to implement $\text{CR}(\theta)$ gates. IBMQ has adopted this method in some of its quantum computers, branding them as fractional gates \cite{IBMQFractionalGates}.

This method is an initial step in what we define as Dynamic Pulse Scheduling. Ideally, the dynamic approach does away with a fixed gate set. Instead, quantum gates are generated at compilation time, tailored to both the quantum circuit and the hardware's characteristics. Rather than creating a universal gate set designed to perform well across all scenarios, the dynamic approach generates specialized pulse schedules based on hardware calibration data, the overall quantum circuit, and the specific gate’s position within the circuit. As a result, the same quantum operation can be implemented using different pulse schedules, depending on the gate’s position.

\cite{Shi2019} follows this idea by proposing a method that aggregates instructions of quantum circuits and generates pulse sequences that implement these aggregated instructions. This approach often leads to much faster pulse sequences compared to the original implementations. However, the compilation time is a limitation of this method, and the evaluation relies on simulations that do not consider quantum hardware restrictions.

Similar to this work, many studies in the literature \cite{Murali2019, Lao2022, Shi2019} also perform optimizations based on a graph model of the quantum circuit, which they typically refer to as the Quantum Operation Dependency Graph. In this work, we go further by exploring techniques from the Critical Path Method (CPM), described in the next section, which facilitates the time management of quantum circuit execution and enables our time-optimization framework.

\section{\label{sec:cpm}Critical Path Method}

The Critical Path Method (CPM) is an algorithm for scheduling a set of activities based on their duration and dependencies \cite{Kelley1959}. It is often used together with the Program Evaluation and Review Technique (PERT), which considers that activities may have variable durations. Here, the diagrams for CPM consider an activity-on-node approach, where activities are represented by nodes and their dependencies by edges.

In CPM, the input is a list of activities with their dependencies and durations. A directed graph is constructed. The objective is to determine, for each activity (node), the Early Start (ES) and Early Finish (EF) times, which represent the earliest possible times an activity can start and finish without delaying the project, and the Late Start (LS) and Late Finish (LF) times, which indicate the latest times an activity can start and finish without affecting the overall schedule.

Afterwards, the critical path is the path that for each of its activities; if one of them takes more time than its minimum duration, the overall duration of the project is increased. In the context of this work, activities are quantum gates, each with a duration dependent on their control pulse. Consequently, within the critical path, quantum gates need to be as fast as possible to minimize circuit duration. Outside the critical path, we can allocate time available based on the ES, EF, LS, and LF times.

The algorithm for finding the times and the critical path consists of two parts, the forward pass and backward pass \cite{Kramer2006}. In forward pass, the ES and EF times and the overall duration are calculated. The algorithm traverses a topological ordering \footnote{A topological sort of a directed and acyclic graph $G=(E,V)$ is a linear ordering of all its vertices such that if $(u,v) \in E$, $u$ appears before $v$ in the ordering \cite{Cormen2022}. The topological sort offers an ordering of the vertices that respects the dependencies indicated by the edges.} defining the ES time as the highest EF value from immediate predecessor nodes and the EF time as the ES plus its duration. The overall duration is EF from the last node. In backward pass, LS and LF times are calculated. The algorithm now traverses an inverse topological ordering, where LF is the minimum of the LS values from immediate successor nodes and LS is the LF minus its duration. Finally, the activities in the critical path are the ones in which ES$=$LS and EF$=$LF. With CPM, the activities in the critical path do not have any tolerance time, they have to be done in their defined duration, otherwise the overall project time will delay. While the activities outside the critical path have a time slack to be performed.

\section{\label{sec:proposal}Time-Optimization Framework}

\begin{figure}
    \centering
    \includegraphics[width=1\linewidth]{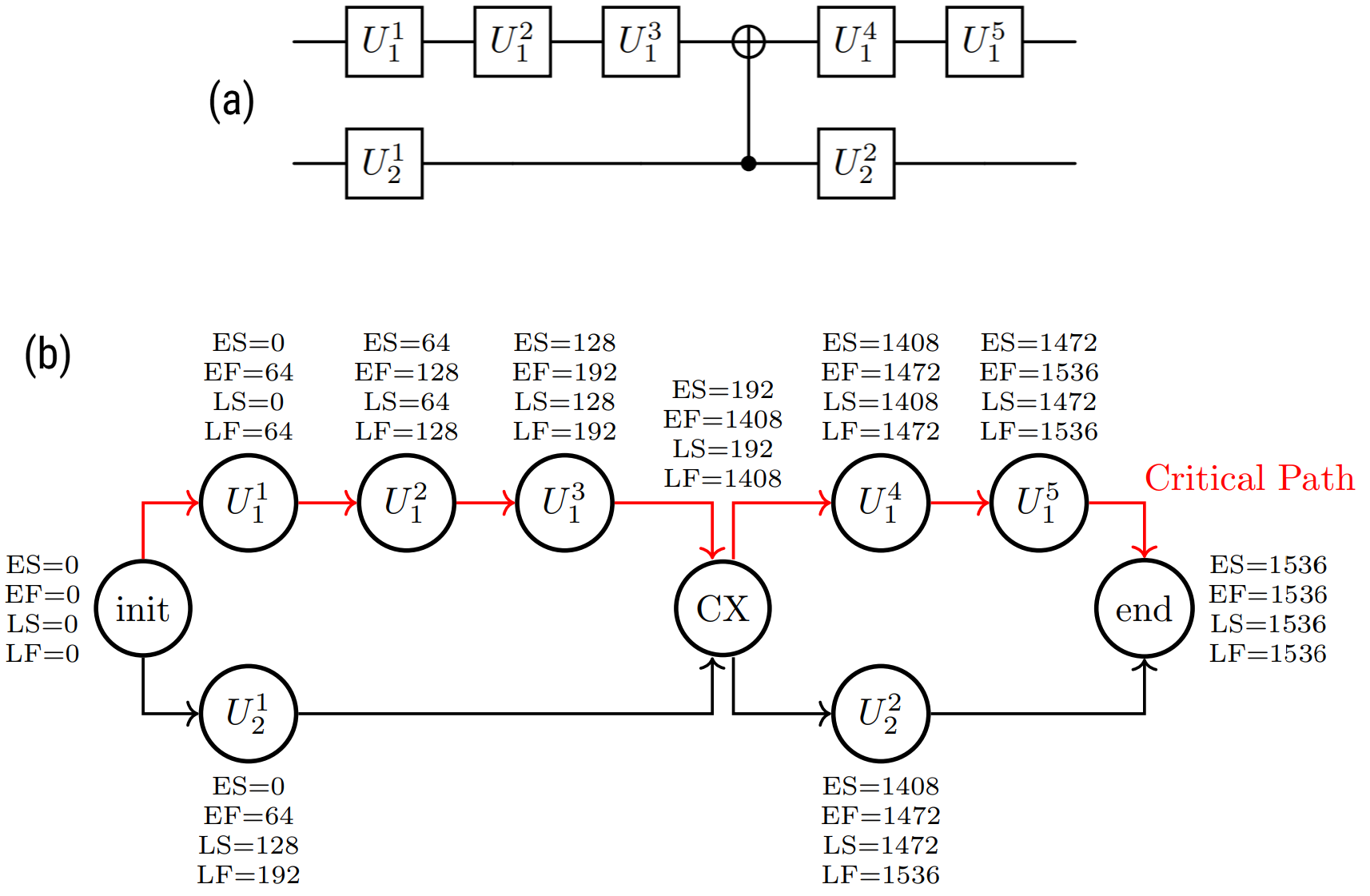}
    \caption{Example of a quantum operation dependency graph construction. (a) An example of a quantum circuit with two idle periods on qubit 2. (b) The quantum operation dependency graph corresponding to the quantum circuit in (a), with the critical path highlighted in red. The ES, EF, LS, and LF times are given in arbitrary units.}
    \label{fig:build-dependency-graph}
\end{figure}

In the static approach, because quantum gates have fixed durations, there may be idle periods in the pulse schedule of multi-qubit quantum circuits. These idle periods occur when $n$ qubits participate in a multi-qubit operation, and certain qubits must wait for others to complete their preceding operations $U_q^i$, where $U_q^i$ represents a generic operation $U$ performed on qubit $q$ (or qubits $q$), and $i$ denotes the operation index. The idle periods are well explored in the context of dynamical decoupling (see \cite{DD}), where an identity sequence of quantum gates is applied to suppress the effects of environmental noise and recover the quantum state. Figure \ref{fig:build-dependency-graph} shows an example of a quantum circuit with two idle periods on the second qubit. Supposing all single-qubit gates have a duration of 64dt, $U_2^1$ and $U_2^2$ have idle periods of 128 and 64dt, respectively. The operations $U_2^1$ and $U_2^2$ can be scheduled at any time of the idle period. For instance, in Qiskit the following schedule policies are applied \cite{QiskitSchedulerPolicies}:
\begin{enumerate}
    \item ``as soon as possible'' applies the operation and idle;
    \item ``as late as possible'' delays applying the operation until the last possible moment.
\end{enumerate}

Alternatively, we follow an ``as long as possible'' policy, where idle periods are used to apply longer and more robust quantum gates. Algorithm \ref{alg:time-opt-framework} presents the overall framework. Given that $n_q$ is the number of qubits, $n_g$ is the number of gates (which corresponds to the number of nodes in the graph), and $n_d$ is the number of allowed durations in the worst case, the algorithm's complexity is polynomial: $\mathcal{O}({n_g}^2n_dn_q)$.

The time-optimization algorithm begins by building the quantum operation dependency graph, as shown in Figure \ref{fig:build-dependency-graph}. The CPM algorithm is executed to define the ES, EF, LS, and LF times for each quantum gate. The gate duration considered by the CPM algorithm is the shortest control pulse calibrated for the operation. The time-optimization algorithm is then executed as presented in Algorithm \ref{alg:time-opt-algorithm}. This algorithm attempts to extend each gate duration based on the CPM, as long as it does not increase the overall circuit duration. The allowed durations for each quantum gate are obtained through the gate set $S$, which might contain different implementations of the same operation $U_q^i$.

\begin{figure}[h]
\begin{algorithm}[H]
    \caption{Time-optimization Framework}
    \label{alg:time-opt-framework}
    \begin{algorithmic}
        \Require $C$: Quantum Circuit, $S$: Gate set
        \State $G \gets$ \Call{create\_graph}{$C$}         \Comment{$\mathcal{O}(n_g)$}
        \State \Call{CPM}{$G$}                             \Comment{$\mathcal{O}\left(n_g(n_q+1)\right)$}
        \State \Call{optimize\_durations}{$G$, $S$}        \Comment{$\mathcal{O}(n_g^2n_dn_q)$}
        \State $P \gets$ \Call{create\_schedule}{$G$, $S$} \Comment{$\mathcal{O}(n_g)$}
        \State \textbf{return} $P$
    \end{algorithmic}
\end{algorithm}
\end{figure}

\begin{figure}[!h]
\begin{algorithm}[H]
    \caption{Time-optimization Algorithm}
    \label{alg:time-opt-algorithm}
    \begin{algorithmic}[1]
        \Require $G$: Quantum operation dependency graph, $S$: Gate set
        \Procedure{optimize\_durations}{$G$, $S$}
            \State $T \gets$\Call{topological\_order}{$G$}  \Comment{$\mathcal{O}(n_g(n_q+1))$}
            \Statex
            \Procedure{update\_CPM}{$g$}    \Comment{$\mathcal{O}(n_gn_q)$}
                \State $i \gets \text{index of } g \text{ in } T$
                \Statex
                \State \texttt{/* Update successors */}
                \For{$u \in T[i:]$}
                    \For{$s \in N^+(u)$}
                        \If{$u\text{.EF} > s\text{.ES}$}
                            \State $s\text{.ES} \gets u\text{.EF}$
                            \State $s\text{.EF} \gets s\text{.ES}+s\text{.duration}$
                        \EndIf
                    \EndFor
                \EndFor
                \Statex
                \State \texttt{/* Update predecessors */}
                \For{$u \in T[i:0]$}    
                    \For{$p \in N^-(u)$}
                        \If{$u\text{.LS} < p\text{.LF}$}
                            \State $p\text{.LF} \gets u\text{.LS}$
                            \State $p\text{.LS} \gets p\text{.LF}-p\text{.duration}$
                        \EndIf
                    \EndFor
                \EndFor
            \EndProcedure
            \Statex
            \State $Q \gets$ \textbf{OrderedQueue()}
            \Statex
            \State \texttt{/* Enqueue gates that are not in critical paths */}
            \For{$g \in V(G)$}  \Comment{$\mathcal{O}(n_g\log n_g)$}
                \If{$g$.ES $\neq$ $g$.LS $\lor$ $g$.EF $\neq$ $g$.LF}
                    \State $Q$.enqueue($g$, $g$.rotation/$g$.duration)
                \EndIf
            \EndFor
            \Statex
            \While{$Q.empty() = \textbf{false}$} \Comment{$\mathcal{O}\left(n_gn_d(n_gn_q+\log n_g)\right)$}
                \State $g \gets Q$.dequeue()
                \State $d \gets $ \Call{get\_next\_duration}{$S$, $g$}
                \If{$g\text{.ES}+d \leq g\text{.LF}$}
                    \State $g\text{.duration} \gets d$
                    \State $g.\text{EF} \gets g\text{.ES}+d$
                    \State $g\text{.LS} \gets g.\text{LF}-d$
                    \State \Call{update\_CPM}{$G, g$}
                    \If{$d <$ \Call{get\_max\_duration}{$S$, $g$}}
                        \State $Q$.enqueue($g$, $g.\text{theta}/g.\text{duration}$)
                    \EndIf
                \EndIf
            \EndWhile 
        \EndProcedure
    \end{algorithmic}
\end{algorithm}
\end{figure}

The time-optimization algorithm (Algorithm \ref{alg:time-opt-algorithm}) works as follows: Each quantum gate might have implementations with variable durations in the gateset $S$. The goal of the algorithm is to increase each gate's duration to its next allowed value, as long as it can finish before its LF time. Since quantum gates implement different rotation angles $\theta$, the algorithm ensures fair opportunities for each gate to increase its duration based on a rate given by $\theta/d$. The justification for this is that operations with larger rotation angles require more time to complete if the same amplitudes are maintained. Therefore, the algorithm uses an ordered queue $Q$ to prioritize gates with larger rotation-to-duration ratios.

The final result of the time-optimization framework is an optimized quantum circuit that maintains the same overall latency as the original, but optimally utilizes idle periods to implement longer and more robust quantum gates. Next, we demonstrate the effectiveness of the algorithm using two approaches for pulse scheduling: static and dynamic.

\section{\label{sec:exp}Experiments and Results}

In this section, we demonstrate the effectiveness of the time optimization framework in a real quantum environment. The experiments were conducted on IBMQ Brisbane using the Qiskit Pulse library. The quantum computer hosts 127 qubits and utilizes the Eagle r3 QPU. The sampling time $dt$ for Brisbane is 0.5 ns. The default pulse durations for this backend are 120$\,dt$ ($\sim$60 ns) for single-qubit gates, and 1320$\,dt$ ($\sim$660 ns) for multi-qubit ECR gates. 

To validate the proposal, we employ Randomized Benchmarking (RB) \cite{RB}. RB is a technique used to assess the fidelity of quantum operations by applying a sequence of random Clifford gates, followed by an inversion gate designed to return the system to its initial state. The fidelity is then estimated by measuring the probability of obtaining the expected outcome ($P(0)$) as a function of the sequence length. Specifically, we generate 50 2-qubit quantum circuits with 1, 41, 81, 121, and 161 Clifford sequences, and 40 3-qubit quantum circuits with 1, 3, 5, and 7 Clifford sequences. 

In this work, only single-qubit gates are addressed, while multi-qubit gates are implemented through the default pulse schedule. For single-qubit gates, Gaussian pulses are employed. Although other pulse shapes may be more suitable in specific situations (e.g., DRAG for faster gates), our focus is on establishing a standardized procedure to evaluate the framework's effectiveness, avoiding calibration variations that may arise from using different pulse shapes.

The only parameter of the Gaussian waveform is the standard deviation, which we determine empirically by:
\begin{equation}
    \sigma(d)=d\times\left(\exp{\left(-\frac{d-68.51}{17.19}\right) + \frac{1}{5}}\right),
\end{equation}
where $d > 17.36$ is the gate duration in $dt$ units. The idea behind this function is to give larger values of $\sigma$ for shorter pulses to avoid amplitude peaks and to approach 20\% of the pulse duration for longer pulses. For example, $\sigma(64)=96$ and $\sigma(120)=30$.

The main difference between static and dynamic experiments is that for a static approach we fine-tuned Gaussian pulses for a Sx gate within a set of durations and use the decomposition presented in equation \ref{eq:static-decomposition}. The allowed durations for the time optimization algorithm are this set of durations for single-qubit gates and the single default duration for the multi-qubit gate ECR. While in the dynamic approach, we use the decomposition in equation \ref{eq:dynamic-decomposition} and implement arbitrary $x$ rotations using interpolated results from default Rabi experiment (see Section \ref{subsec:rabi-exp}). So in this approach, any discrete value in a certain interval is possible for single-qubit pulse duration, but the pulse duration must be a multiple of 8 $dt$ in Brisbane quantum computer. For both approaches, we compare fixed-duration gates to variable-duration gates determined by the time-optimization algorithm.

\subsection{\label{subsec:static}Static Pulse Scheduling}

In static approach, Gaussian control pulses with durations of $32$, $48$, $64$, $120$, $256$, and $512\,dt$ were calibrated and fine-tuned for each qubit. The experiments compare fixed-duration quantum gates with time-optimized gates. Specifically, we compare fixed-duration gates of $32$, $64$, and $120\,dt$ to time-optimized gates allowing durations within the ranges $32–512$, $64–512$, and $120–512\,dt$.

\begin{figure}
    \centering
    \includegraphics[width=1\linewidth]{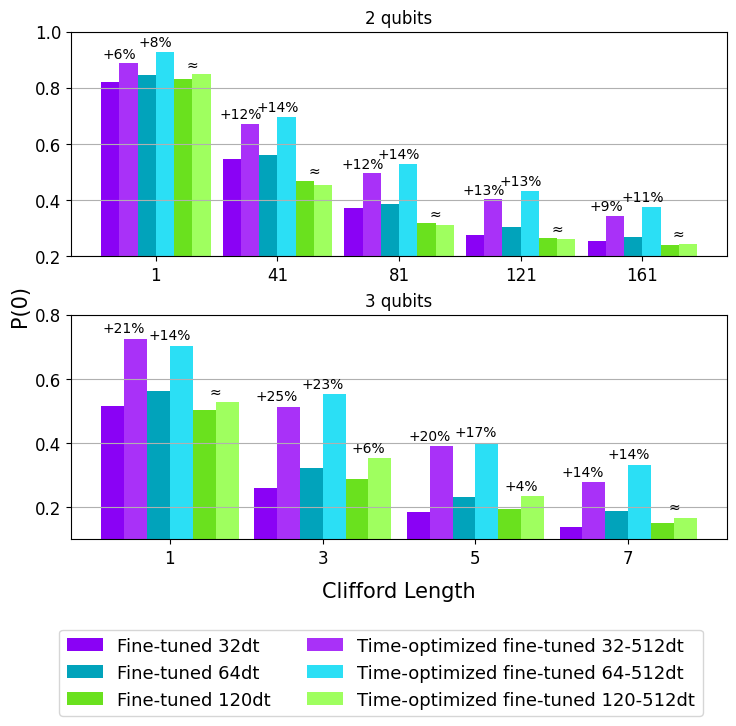}
    \caption{Randomized benchmarking results for static fine-tuned gates. Each bar represents the mean success probability of 10 RB quantum circuits. The success probability $P(0)$ represents the likelihood of measuring the system in the initial state, which is used to evaluate gate fidelity through RB. The Clifford length, shown on the x-axis, represents the number of Clifford layers in the sequence, where each layer corresponds to a random multi-qubit Clifford operation decomposed into a specific number of single- and multi-qubit gates.}
    \label{fig:static}
\end{figure}

Figure \ref{fig:static} shows the results obtained from the randomized benchmarking. In the vast majority of cases tested, the time-optimized quantum circuits outperformed the fixed-duration ones. Gains of more than 20\% in absolute success probability were observed for 3-qubit quantum circuits, with some cases showing improvements exceeding 25\%. The greatest advantages of using the algorithm were seen in the most extreme scenarios: quantum circuits with more qubits and faster quantum gates.

The main challenge of using faster Gaussian pulses is leakage. The amplitude peak around the center of the Gaussian waveform can be strong enough to excite states beyond $\ket{1}$ and cause other errors. On the other hand, the main issue with longer quantum gates is the extended circuit latency and the effects of decoherence. Therefore, it is expected that the time-optimization framework applied to gate sets with longer quantum gates does not provide significant advantages, as the goal of the algorithm is precisely to maintain low latency while avoiding errors from fast quantum gates, which are only retained in critical paths. However, this technique is highly beneficial when faster quantum gates are present in the gate set.

Additionally, as the number of qubits in the quantum circuits increases, the number and duration of idle periods are also expected to increase. Consequently, it is expected that the algorithm will have more space to implement longer and more robust quantum gates, while the number of fast, noisier quantum gates will become increasingly smaller in proportion to the longer ones, which will become the majority. While our experiments were limited to circuits with up to three qubits, a comparison of the results from 2-qubit and 3-qubit circuits suggests that even greater advantages would be observed if the technique were applied to circuits with more qubits.

\begin{figure}
    \centering
    \includegraphics[width=1\linewidth]{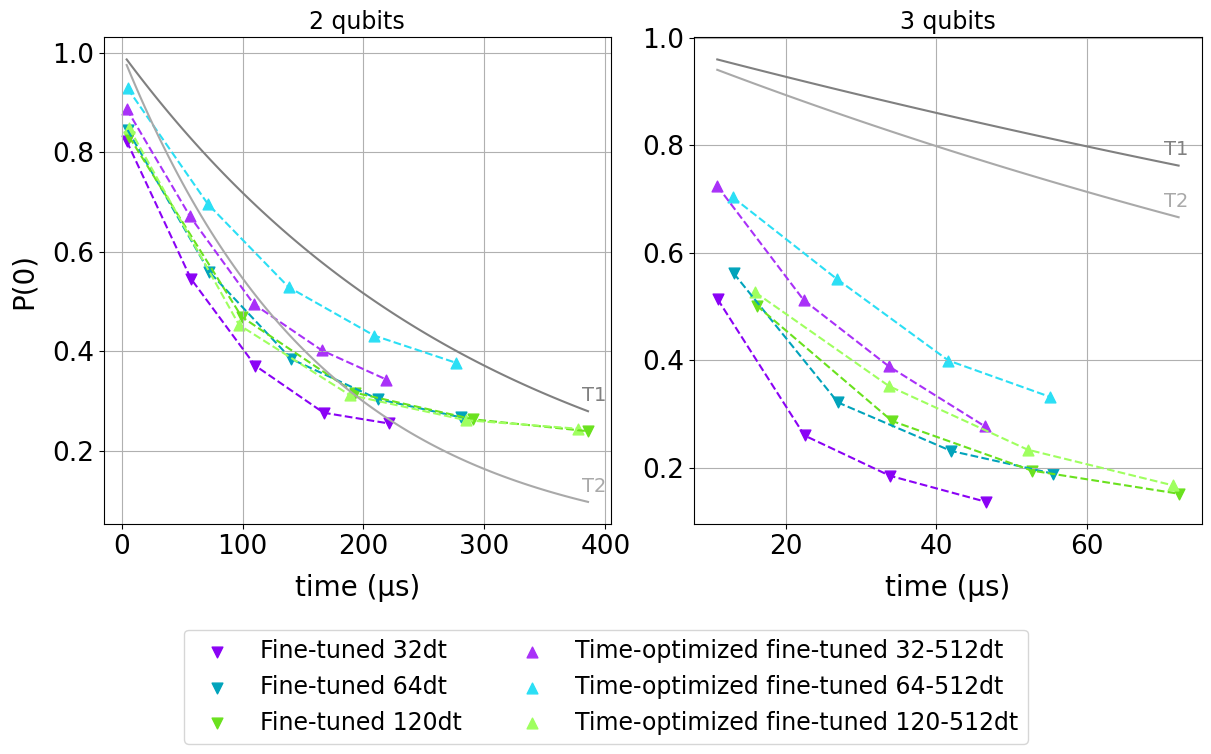}
    \caption{Randomized benchmarking results for static fine-tuned gates in time-scale. The experimental results ($P(0)$), circuit latencies, and decoherence times are averaged. Down-pointing triangles represent the mean $P(0)$ for the default fixed-duration gates, while up-pointing triangles represent the mean $P(0)$ for the pulse scheduling under the time-optimization framework. The $T_1$ and $T_2$ times were obtained from IBMQ jobs and plotted using an exponential decay function $e^{-t/T}$. For the $T_1$ plot, consider the decay from $\ket{1}$ to $\ket{0}$, or equivalently, $P(1)$ on the y-axis.}
    \label{fig:static-timescale}
\end{figure}

Figure \ref{fig:static-timescale} presents the results on a timescale. The decoherence times shown are averaged across the qubits involved and all executions represented in each graphic. Two key observations emerge from this perspective. First, the proximity of the computation times to the decoherence limits is more evident. Time-optimized circuits are consistently closer to the decoherence limits. However, when comparing quantum circuits with varying latencies, it is important to consider that lower-latency circuits complete their computations earlier. As a result, they can achieve better outcomes despite being farther from the decoherence limits. For instance, even though the fine-tuned $32\,dt$ line is below the fine-tuned $120\,dt$ line, the earlier completion of the $32\,dt$ circuit leads to final results that are closer to the expected values. 

When observing the decoherence limits, time-optimized $32\,dt$ and $64\,dt$ circuits approach the decoherence limits most closely. Nevertheless, the primary limitation for 3-qubit quantum circuits lies not in decoherence but in gate quality. Even though, these circuits demonstrate the most significant benefits from the time-optimization framework.

\subsection{\label{subsec:dynamic}Dynamic Pulse Scheduling}

In the dynamic approach, the time-optimization framework was benchmarked by comparing default dynamic Gaussian pulses of $32$, $48$, and $64\,dt$ durations, with a maximum duration of up to $128\,dt$ (for $\pi/2$ rotations). In this approach, the algorithm can select any integer duration multiple of 8 within this range for each quantum gate. Additionally, gate durations are normalized based on the operation they implement. For instance, if the minimum duration for a $\pi/2$ rotation is set to $32\,dt$, the minimum duration for a $\pi$ rotation would be $64\,dt$.

There are specific reasons for defining a maximum duration. First, a duration of $512\,dt$ is generally sufficient to implement noise-robust single-qubit gates. Second, IBMQ imposes a limitation on the number of pulse samples that can be included within a single job, making it impractical to perform Rabi experiments for excessively small amplitude values associated with longer quantum gates. This limitation is particularly significant in the dynamic approach, which relies heavily on simple Rabi experiment. Therefore, in this context, a maximum duration of $128\,dt$ for $\pi/2$ rotations was chosen as appropriate for a Rabi experiment with a minimum amplitude value of 0.001, yielding a Rabi frequency of approximately 105 kHz (for $4\pi$). Additionally, reducing the maximum duration can sometimes yield better results, as the extra idle periods may be utilized by other quantum gates.

\begin{figure}
    \centering
    \includegraphics[width=1\linewidth]{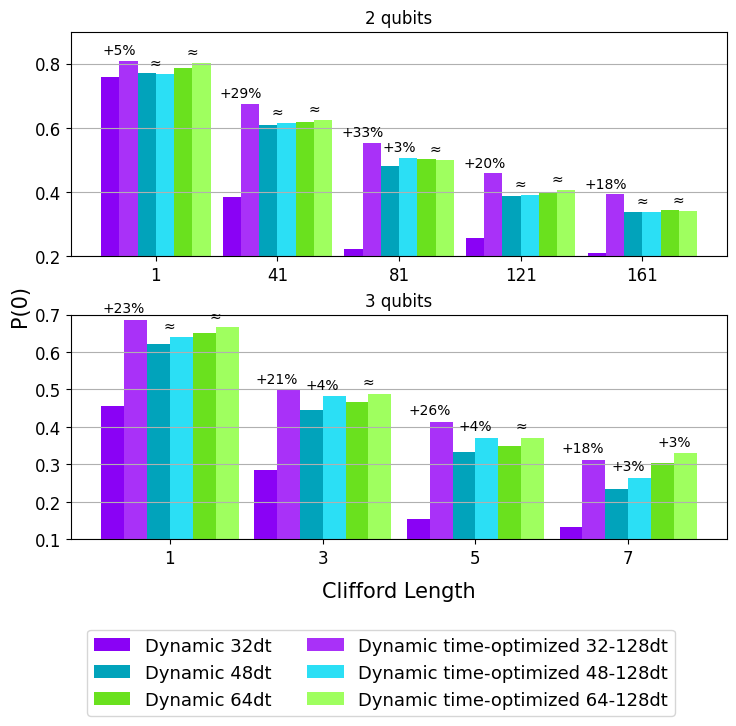}
    \caption{Randomized benchmarking results for dynamic quantum gates. Each bar represents the mean success probability of 10 RB quantum circuits. The duration limits are for $\pi/2$ rotations. The success probability $P(0)$ represents the likelihood of measuring the system in the initial state, which is used to evaluate gate fidelity through RB. The Clifford length, shown on the x-axis, represents the number of Clifford layers in the sequence, where each layer corresponds to a random multi-qubit Clifford operation decomposed into a specific number of single- and multi-qubit gates.}
    \label{fig:dynamic}
\end{figure}

Figure \ref{fig:dynamic} presents the results for the dynamic gate generation. While improvements exceeding 30\% are observed for the $32\,dt$ experiments, the results differ significantly from those of the static approach. For fixed-duration dynamic gates, the $32\,dt$ experiments perform much worse than the $48\,dt$ ones, which are only slightly worse than $64\,dt$. However, in time-optimized quantum circuits with a minimum duration of $32\,dt$, better performance is observed in most cases. Furthermore, this scenario demonstrates more significant improvements compared to others. For $48\,dt$ and $64\,dt$ experiments, only slight improvements are achieved.

Although these results may appear counterintuitive at first, they are logically consistent. In the static approach, longer quantum gates are meticulously fine-tuned, resulting in higher fidelity. Conversely, in the dynamic approach, longer quantum gates do not necessarily have higher fidelity because they are not fine-tuned, so it is hard to ensure consistent gate fidelity across varying durations. Consequently, better results are achieved when quantum gates are fast enough to introduce errors, which are then suppressed by the algorithm. The low latency compensates for noisy pulses on the critical path, leading to improved performance. However, it is important to emphasize that results may vary across different quantum computers. Nonetheless, the time-optimization framework demonstrates improvements in most scenarios.

\begin{figure}
    \centering
    \includegraphics[width=1\linewidth]{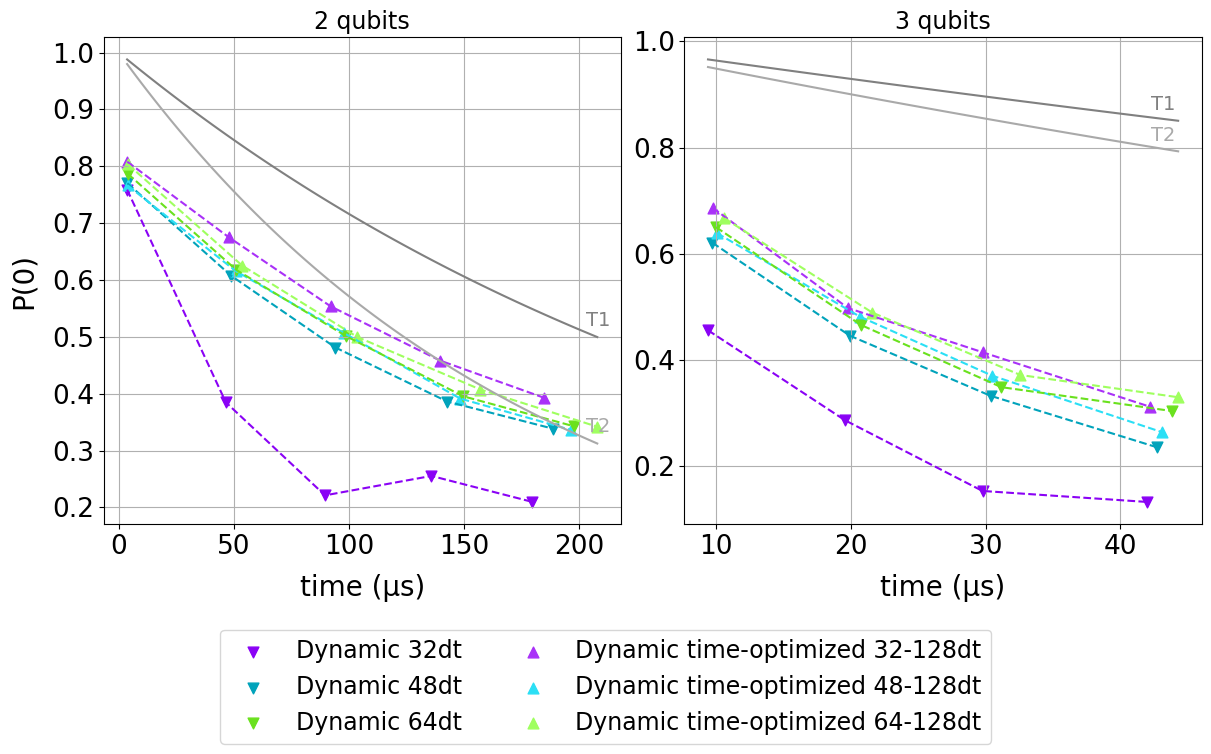}
    \caption{Randomized benchmarking results for dynamic quantum gates in time-scale. The duration limits are for $\pi/2$ rotation. The results, circuit latencies and decoherence times are averaged. Down-pointing triangles represent the mean $P(0)$ for the default fixed-duration gates, while up-pointing triangles represent the mean $P(0)$ for the pulse scheduling under the time-optimization framework. The $T_1$ and $T_2$ times were obtained from IBMQ jobs and plotted using an exponential decay function $e^{-t/T}$. For the $T_1$ plot, consider the decay from $\ket{1}$ to $\ket{0}$, or equivalently, $P(1)$ on the y-axis.}
    \label{fig:dynamic-timescale}
\end{figure}

Furthermore, Figure \ref{fig:dynamic-timescale} presents the same results in the time domain. This perspective makes it clearer that the time-optimization lines are above the fixed-duration ones across all cases, particularly for $32\,dt$, which showed significantly improved results. Similar trends to the static approach are observed when analyzing the decoherence limits. For 2-qubit circuits, the results are closer to the decoherence limits, whereas for 3-qubit circuits, decoherence does not seems to be the primary performance constraint. Nevertheless, quantum circuits in the static approach demonstrate the ability to achieve results closer to the decoherence limits.

\subsection{Timing Analysis}

\begin{figure}
    \centering
    \includegraphics[width=1\linewidth]{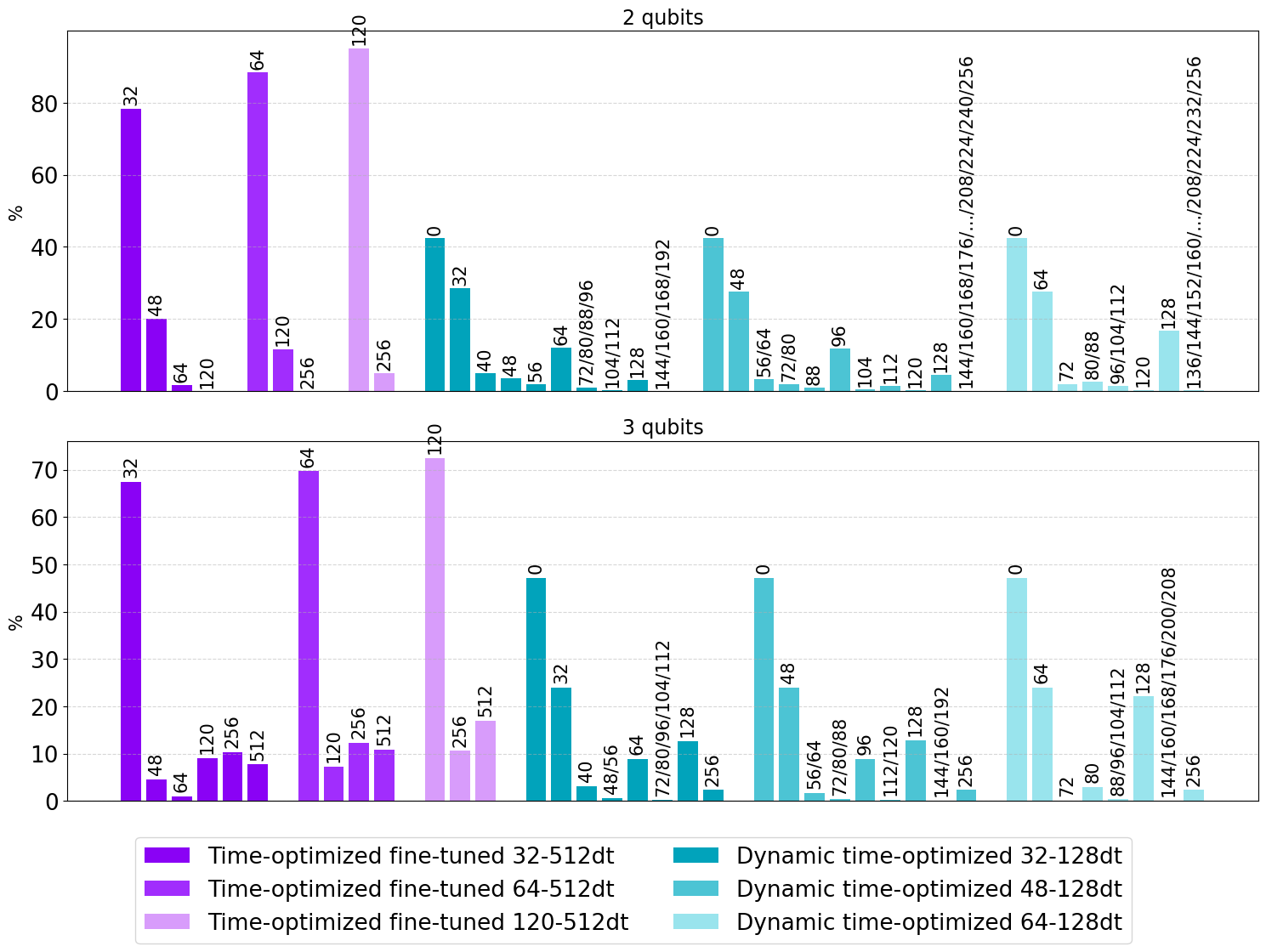}
    \caption{Pulse duration distribution from the time-optimization algorithm for static and dynamic approaches. Results are averaged over subsequence durations with similar frequency values.}
    \label{fig:duration-frequency-distribution}
\end{figure}

An important aspect to analyze are the durations selected by the time-optimization algorithm. Figure \ref{fig:duration-frequency-distribution} shows the frequency distribution of pulse durations for both static and dynamic approaches.

First, some clarifications are necessary. In the dynamic approach, single-qubit quantum gates are implemented according to Equation \ref{eq:dynamic-decomposition}. If $\theta=0$, the quantum gate consists only of virtual Z gates and has no duration. Additionally, in the dynamic approach, the duration units correspond to $\pi/2$ rotations. For instance, a duration range of $32-128\,dt$ is for $\pi/2$ rotations, while for $\pi$ rotations, the equivalent range is $64-256\,dt$. Using the Qiskit transpiler, the $\theta$ values for U3 gates are restricted to $0$, $\pi$, and $\pi/2$ angles. Therefore, in this example, the duration interval spans from $0$ to $256\,dt$.

From the experimental results, it was observed that in the static approach, quantum circuits with more qubits achieved the greatest improvements using the time-optimization framework. In the dynamic approach, significant improvements were limited to experiments with durations of $48$ and $64\,dt$. Figure \ref{fig:duration-frequency-distribution} shows that, in static approach, the percentage of quantum gates with minimum duration decreases from 78\%, 88\%, and 95\% for 2-qubit circuits to 67\%, 70\%, and 72\% for 3-qubit circuits. In the dynamic approach, due to the broader range of possible gate durations, the frequency of faster quantum gates is already low for 2-qubit circuits and decreases slightly for 3-qubit circuits. Furthermore, longer gate durations are more prevalent in 3-qubit circuits. For example, in the static approach, gates with a duration of $512\,dt$ appear exclusively in 3-qubit circuits.

In that regard, we expect that experiments with more qubits would benefit even further from the proposed framework. When utilizing all 127 qubits of Brisbane, only a small subset of quantum gates on the critical path would be implemented as fast quantum gates, which will focus on reducing their intrinsic leakage errors. The remaining quantum gates, which would constitute the majority, could be implemented using noise-robust waveforms. Additionally, since the time-optimization algorithm guarantees no increase in latency, quantum gate generation methods do not necessarily need to account for gate duration constraints. Furthermore, the algorithm's complexity is polynomial, making it feasible to execute during compilation time.

\section{\label{sec:conclusion}Conclusion and Future Works}

In this paper, we proposed a framework that allows the use of noise-robust quantum gates with longer durations without compromising quantum circuit latency. By applying techniques from project management and the critical path method, we developed an algorithm that selects optimal quantum gate implementations from a gate set. The algorithm places fast quantum gates on the critical path and utilizes idle periods to implement longer, more robust gates. We demonstrated that this algorithm runs in polynomial time and can be efficiently executed during compilation.

Experiments conducted on IBMQ Brisbane demonstrated the effectiveness of this framework. By employing Gaussian control pulses with varying durations, we were able to execute quantum circuits with significantly reduced latency without being restricted to the susceptibility to leakage inherent in fast control pulses and the inability to incorporate robustness on them. The improvements achieved exceed 25\% in the absolute success probability of quantum circuit execution, with even greater gains expected for circuits involving more qubits.

Due to restricted access to quantum hardware, we were unable to scale the experiments to real-world scenarios. In future work, we aim to conduct experiments in practical scenarios, utilizing more qubits and more sophisticated control pulses. The use of Gaussian control pulses with varying durations does not represent the best scenario for the framework, as robustness is not actually incorporated into the waveform of longer gates. Therefore, future experiments could incorporate DRAG and related techniques \cite{DRAG, FASTDRAG} as well as optimization-based waveforms \cite{Carvalho2021}.

Additionally, we believe significant attention should be given to the specialization of quantum gates. Currently, only a single gate implementation is typically calibrated for each operation. However, it is quite difficult to calibrate a single waveform that performs well in all scenarios. We propose that calibrating multiple waveforms for the same operation could offer substantial advantages for quantum computing. Nevertheless, other variables beyond gate duration could be explored. In future work, we intend to investigate how different quantum errors behave in the quantum circuit with the objective to incorporate waveform selection also based on their robustness to different error sources. 

Ultimately, this specialization of quantum gates leads to dynamic gate generation, where waveforms are generated at compilation time based on detailed hardware characterization. Despite the potential of AI-based techniques, significant advancements are still required to fully realize the dynamic approach.

\begin{acknowledgments}
This work was supported by Conselho Nacional de Desenvolvimento Científico e Tecnológico - CNPq through grant number 409673/2022-6.
\end{acknowledgments}




\bibliography{apssamp}

\end{document}